\documentclass[preprintnumbers, aps, floatfix, onecolumn,
preprintnumbers, letterpaper, superscriptaddress,nofootinbib,11pt]{revtex4}
\usepackage{amsmath}
\usepackage{hyperref}
\usepackage{amsfonts}
\usepackage{amssymb}
\usepackage{graphicx}
\usepackage{latexsym}
\usepackage{color}
\usepackage{booktabs}
\usepackage{dcolumn}
\usepackage{epsfig}
\usepackage{subfigure}
\usepackage{float}
\usepackage{multirow}
\usepackage{ulem}
\usepackage{xcolor}
\usepackage{epstopdf}

\def\be#1\ee{\begin{align}#1\end{align}} 
\renewcommand{\Re}{\operatorname{Re}}
\renewcommand{\Im}{\operatorname{Im}}
\def\be{\begin{eqnarray}}
\def\ee{\end{eqnarray}}

\begin{document}

\title{Instability of Reissner-Nordstr\"om-AdS black hole under perturbations of a scalar field
 coupled to Einstein tensor.}

 \author{E. Abdalla}\email{eabdalla@usp.br}
 \affiliation{Instituto de F\'isica, Universidade de S\~ao Paulo,
   Caixa Postal 66318, CEP 05314-970, S\~ao Paulo, SP, Brazil}
 \author{B. Cuadros-Melgar}\email{bertha@usp.br}
 \affiliation{Escola de Engenharia de Lorena, Universidade de S\~ao
   Paulo, Estrada Municipal do Campinho S/N, CEP 12602-810, Lorena, SP, Brazil}
 \author{R. D. B. Fontana}\email{rodrigo.fontana@uffs.edu.br}
 \affiliation{Universidade Federal da Fronteira Sul, Campus Chapec\'o, CEP 89802-112, SC, Brazil}
 \author{Jeferson de Oliveira}\email{jeferson@gravitacao.org}
 \affiliation{Instituto de F\'i­sica, Universidade Federal de Mato Grosso, CEP 78060-900, Cuiab\'a, MT, Brazil}
\author{Eleftherios Papantonopoulos}\email{lpapa@central.ntua.gr}
 \affiliation{Department of Physics, National Technical University of Athens, Zografou Campus GR 157 73, Athens, Greece}
\author{A. B. Pavan}\email{alanbpavan@gmail.com}
 \affiliation{Universidade Federal de Itajub\'a, Instituto de F\'isica e Qu\'{\i}mica, CEP 37500-903, Itajub\'a, MG, Brazil}

\date{\today}
\begin{abstract}

We study the instability of a  Reissner-Nordstr\"om-AdS (RNAdS) black hole under perturbations of a massive
scalar field coupled to Einstein tensor. Calculating the potential of the scalar perturbations
 we find that as the strength of the coupling of the scalar to Einstein tensor is increasing, the potential develops a negative well
outside the black hole horizon, indicating an instability of the background RNAdS. We then investigate the effect of this coupling on the
quasinormal modes. We find that there exists a critical value of the coupling which triggers the instability of the RNAdS.
We also find that as the charge of the RNAdS is increased towards its extremal value, the critical value of the  derivative coupling
is decreased.

\end{abstract}

\maketitle

\section{Introduction}\label{intro}

Recently, there has been an intense interest in the study of gravitational theories that modify the Einstein's theory of gravity. One  class of these theories concern the scalar-tensor theories.  The most well studied
scalar-tensor models  are those described by the Horndeski Lagrangian
\cite{horny} which gives second order field equations in four
dimensions \cite{Nicolis:2008in,Deffayet:2009wt,Deffayet:2009mn}. One important term appearing in the Horndeski Lagrangian is
the kinetic coupling  of a  scalar field to curvature. This derivative coupling of matter to gravity
has interesting cosmological implications, acting as a friction term in the early inflationary cosmological evolution \cite{Amendola:1993uh,Sushkov:2009hk,Saridakis:2010mf,Granda:2012tx,Germani:2010gm}. Observational tests of inflation with a field coupled to Einstein
tensor were presented in \cite{Tsujikawa:2012mk}. It has also been shown that the derivative coupling to gravity provides a natural mechanism to suppress the overproduction of heavy particles after inflation \cite{Koutsoumbas:2013boa}. Particle
production after  the end of inflation in the presence of the
derivative coupling was also discussed in \cite{Sadjadi:2012zp}.

This interesting cosmological behaviour  of the coupling of the scalar field to Einstein tensor is a consequence of the fact
 that this term introduces a scale in the theory and
effectively acts as a
 cosmological constant \cite{Sushkov:2009hk}. Thus,  the presence of the cosmological constant alters the local
 properties of the spacetime allowing in this way  the generation of hairy black hole solutions.
 However, in \cite{Hui:2012qt} it was shown that in Galileon theories there are stringent constraints that these solutions  have to respect in order  the
scalar field  to have non-trivial profile and to be finite on the horizon.

 One of the first black hole solutions with derivative coupling \cite{Rinaldi:2012vy} failed to  evade singular behaviour and  the scalar field blows up  on the horizon. There are ways to evade this problem and one of them is to
 break the shift symmetry of the scalar field by introducing a mass term for
the scalar field \cite{Kolyvaris:2011fk,Kolyvaris:2013zfa}. Another way is to  allow the scalar field to be time dependent, while keeping the shift symmetry \cite{Babichev:2013cya}.
This permits asymptotically flat (or de-Sitter) solutions and
it gives regular hairy black hole solutions. Then, various  black hole solutions appeared in the literature \cite{Charmousis:2014zaa,Anabalon:2013oea,Babichev:2015rva,Sotiriou:2014pfa,Benkel:2016rlz}.

The stability of gravity theories in the presence of the derivative coupling has been studied as well.  Calculating the quasinormal
spectrum of scalar perturbations in a gravity model with a scalar
field coupled to Einstein tensor an instability was found outside
the horizon of a Reissner-Nordstr\"om black hole
\cite{Chen:2010qf}. It was shown  that for higher angular momentum
and  for large values of the derivative coupling, the effective
potential develops a negative gap near the
black hole horizon. This can be interpreted as a signal that a
 phase transition has occurred  to a hairy black hole configuration.

 This effect was further investigated in  \cite{Kolyvaris:2011fk}. Keeping a vanishing cosmological constant and no derivative coupling,
 introducing an electromagnetic field, we do not have the so called "geometrical" breaking
of the Abelian symmetry near the black hole horizon  \cite{Gubser:2008px}. However, turning on the derivative coupling, it was shown in \cite{Kolyvaris:2011fk} that
 there is a critical temperature below which there is a phase transition  to a
 hairy black hole configuration. This is happening because the space is
asymptotically  quasi-AdS, due to the presence of the derivative coupling, and a new hairy black hole configuration is generated as the result of the breaking
 of an Abelian gauge symmetry by curvature effects. It was also found that this hairy black hole
 configuration is spherically symmetric and it is
 thermodynamically stable, having larger temperature than the corresponding  Reissner-Nordstr\"om black
hole.

The  QNMs of a test massless scalar field coupled to Einstein tensor were calculated in \cite{Minamitsuji:2014hha}. Also the QNMs were studied in \cite{Yu:2018zqd} for various static and spherically symmetric black holes in the presence of the derivative coupling. It was found that the oscillation
of QNMs becomes slower and slower and the decay of QNMs becomes faster and faster  with the increasing of the derivative coupling confirming in this way the findings that the coupling of the scalar field to curvature changes the kinetic properties of the scalar field influencing  the decay of the QNMs.
 Calculations of QNMs
for a massive scalar field
with the derivative coupling in the background of  Reissner-Nordstr\"om black hole
were performed in \cite{Konoplya:2018qov}.

The  vectorial and spinorial perturbations were performed in Galileon black holes and the QNMs were
calculated in the presence of the derivative coupling \cite{Abdalla:2018ggo}. The  effect  of  the  derivative  coupling  in  the  quasinormal  spectrum  has  been analysed and evaluated. No instability was found under both vectorial and spinorial perturbations. Also the
  superradiant instability of Galileon black holes was studied in  \cite{Kolyvaris:2017efz,Kolyvaris:2018zxl}. A massive charged scalar wave coupled   to curvature was scattered off the horizon of a  Horndeski black hole. It was found that a trapping potential is formed outside the horizon of a Horndeski black hole leading to the instability of the Horndeski black hole,  and  the superradiance condition  was calculated.  Also the bound states trapped in the potential well or penetrating the horizon of the Galileon black hole leading to its instability were calculated in \cite{Koutsoumbas:2018gbd}. We also found quasiresonant modes, that is, long lived modes, for fermionic perturbations.

In the present work we study possible instabilities of a  Reissner-Nordstr\"om-AdS black hole by calculating the QNMs of scalar perturbations of a scalar field coupled to Einstein tensor. In an AdS space there is a natural boundary defined by its length $L$ on which the scalar wave is scattered back. As we already discussed, the derivative coupling introduces another scale and one of the aims of this work is to study the interplay of these scales and their effects on the stability of the background Reissner-Nordstr\"om-AdS black hole. We will also investigate the resonant transfer of energy from low to high frequencies because we expect that at these regimes  this transfer of energy results  to the instability of the Reissner-Nordstr\"om-AdS black hole and we will find the critical value of the derivative coupling at which this behaviour occurs. Finally, we will investigate what is the effect
of the behaviour of the derivative coupling on the QNMs to alter the kinetic properties of the scalar field.

The work is organized as follows. In Section \ref{sec1} we consider a
massive scalar field coupled to Einstein tensor propagating on a fixed
AdS background. In Section \ref{sec2} we  study the potential formed
outside the event horizon of
Reissner-Nordstr\"om-AdS black holes and possible instabilities. In
Section \ref{sec3} we calculate the QNMs and finally in Section
\ref{sec5} we discuss our conclusions.

\section{Massive scalar field coupled to Einstein tensor}\label{sec1}

We consider the evolution of a massive scalar field $\Phi$ coupled to
Einstein tensor propagating in AdS geometries. We consider a
massive scalar field interacting with the Einstein tensor
$G_{\mu\nu}$ \cite{Amendola:1993uh,Sushkov:2009hk}
as
\be
\label{e1}
{\cal L}_{pert} = -\frac{\sqrt{-g}}{2}\left[\left(g^{\mu\nu}-\eta
G^{\mu\nu}\right) \partial_{\mu}\Phi\partial_{\nu}\Phi+ m^2 \Phi^2 \right]~,
\ee
where $\eta$ is the non-minimally derivative coupling parameter and $m$ is the scalar field mass. More specifically, we investigate the propagation of this scalar field
in spherically symmetric background black hole solutions,  that is,
\be
\label{e2}
&&ds^2 = -F(r)\ dt^2+ F(r)^{-1}\ dr^2+r^2\ d\Omega^2~,
\ee
where $d\Omega^2=d\theta^2+\sin^2\theta d\varphi^2$ is the 2-sphere line
element.

The equation of motion for the scalar field $\Phi$ derived from the Lagrangian (\ref{e1}) can be put in the form
\be
\label{e6}
\frac{1}{\sqrt{-g}}\partial_{\mu}\left(\frac{}{}\sqrt{-g}\ h^{\mu\nu}\ \partial_{\nu}\Phi\right)-m^2 \Phi=0~,
\ee
where $g$ is the determinant of the metric given by the line element (\ref{e2}) and $h^{\mu\nu}$ acts as an ``effective'' metric, and is given by
\begin{eqnarray}
\label{e7}
h^{\mu\nu}=g^{\mu\nu}-\eta\ G^{\mu\nu}~.
\end{eqnarray}
Using the spherically symmetric background given by Eq.(\ref{e2}) we can rewrite Eq.(\ref{e6}) as
\begin{eqnarray}
\label{e10}
-\ddot{\Phi}+ F(r)^2 \Phi''+\ S(r)\ \Phi'- \vartheta (r) \Phi=0~,
\end{eqnarray}
where the dot corresponds to a time derivative, prime corresponds to a radial derivative, and
\begin{eqnarray}
\label{e11}
\vartheta (r)&=&\frac{F}{(1+\eta A)}\left( \frac{l(l+1)}{r^2}(1-\eta B)+m^2\right)~,\\
\nonumber\\
\label{e12}
S(r)&=&F^2 \left(\frac{\eta A'}{1+\eta A}+\frac{F'}{F}+\frac{2}{r}\right)~,
\end{eqnarray}
with functions $A(r)$ and $B(r)$ given by
\begin{eqnarray}\label{AB}
A(r)&=&\left(-\frac{F' }{ r}+\frac{1-F}{r^2}\right)~,\\
B(r)&=& A(r)-\frac{1}{2}\mathcal{R}~,
\end{eqnarray}
where $\mathcal{R}$ is the Ricci scalar corresponding to metric (\ref{e2}).


Introducing the tortoise coordinate  $dr_*=\frac{1}{F}dr$ and considering the separation of variables for the scalar field as given by
\begin{eqnarray}
\label{e13}
\Phi(t,r,\theta,\varphi)=\sum_{l ,\mathfrak{m}}\frac{Z(r,t)}{r (1+\eta A)^{1/2}} Y_{l ,\mathfrak{m}}(\theta,\varphi)~,
\end{eqnarray}
where $Y_{l ,\mathfrak{m}}(\theta,\varphi)$ are the well-known spherical harmonics, the scalar field equation (\ref{e10}) becomes
\be
\label{e14}
-\frac{\partial^2 Z}{\partial t^2} + \frac{\partial^2 Z}{\partial r_{*}^2}  - V_{s}(r) Z = 0~,
\ee
where
\begin{eqnarray}
\label{e15}
V_s(r)&=&\frac{F}{(1+\eta A)} \left[\frac{l(l+1)}{r^2}(1-\eta B)+m^2 +  \frac{F'}{r}(1+\eta A) \right]+F^2 V_{\eta}(r)\,, \\
\label{e16}
V_\eta (r)&=& \frac{\eta}{1+\eta A} \left( \frac{A''}{2} +
\frac{A'F'}{2F}+\frac{A'}{r} \right) - \frac{1}{4} \left(\frac{\eta
  A'}{1+\eta A}\right)^2 \, ,
\end{eqnarray}
with $l$ standing for the scalar field multipole number.

\section{Fixing the background: AdS black holes}\label{sec2}

In this Section we mainly study the propagation of the scalar field in the
Reissner-Nordstr\"om-AdS black hole background and we comment  for the case of Schwarzschild-AdS
in the presence of the derivative coupling, as discussed in the
previous Section.  The equation of motion follows exactly the form of
Eqs.(\ref{e10}), (\ref{e15}), and (\ref{e16}), where the
line-element function $F$ is given by
\be
\label{h1}
F=1-\frac{2M}{r}+\frac{Q^2}{r^2} + \frac{r^2}{L^2}~,
\ee
and the potential functions $A$ and $B$ become
\be
\label{h2}
A=-\frac{3}{L^2}+\frac{Q^2}{r^4}, \hspace{1.5cm} B=\frac{3}{L^2}+\frac{Q^2}{r^4}~,
\ee
for black holes with mass $M$, charge $Q$, and AdS radius $L$. In the
Schwarzschild-AdS black hole case the Klein-Gordon potential will not be affected by the derivative coupling $\eta$ except for the case of massive scalar field.

For the Reissner-Nordstr\"om AdS black hole, we choose  $M$, $Q$, and $L$ such that two horizons are present, the event horizon $r_h$ and the Cauchy horizon $r_c$, in order to prevent naked singularities. This condition in general represents, given the values of  $M$ and $L$, a maximum value for the charge of the black hole, $Q_{ext}$, which is the positive real root of the equation
\be
\label{h3}
\frac{36M^2Q^2-27M^4-8Q^4}{M^2-Q^2} + \sqrt{\frac{729M^8-1944M^6Q^2+1728M^4Q^4-512M^2Q^6}{(M^2-Q^2)^2}}=2L^2.
\ee
The field transformation introduced in Eq. (\ref{e13}) results in a
drawback for its numerical evolution. No matter which method is used, for certain range of parameters of the black hole and coupling, we have a discontinuity in the potential at a specific value of $r\equiv r_d$, defined as
\be
\label{h4}
r_d= \left( \frac{\eta Q^2}{\frac{3\eta}{L^2}-1}\right)^{1/4}~.
\ee
By limiting our investigation to the region beyond the event horizon
and spatial infinity, however, we can prevent such instability to occur in the field equation choosing the range of parameters for which $r_d < r_h$.

The potential presented in Section II depends on four different
parameters, the multipole number $l$, the derivative coupling
parameter $\eta$, the perturbations mass $m$, and the black hole
charge $Q$~\footnote{The parameters we choose to fix are the black
  hole mass $M$ and the AdS radius $L$. Thus, the event horizon will
  only depend on the charge value.}.
We will discuss the behaviour of this potential outside the event horizon and possible instabilities generated by the incident wave depending on the range of accepted parameters.

Graphically analising the potential in Eq.(\ref{e15}) we notice some
general features.
For $Q\not= 0$ (Reissner-Nordstr\"om-AdS case) the potential develops
a negative well whose width, depth, and position depend on the black
hole and perturbation parameters. After this well the potential
becomes positive definite, it develops a local maximum
(peak) and goes to infinity for large $r$, as expected. Due to the
appearance of a negative potential in a certain region we can foresee the
possibility of finding instabilities for some range of parameters.

\begin{figure}[!ht]
\begin{center}
\epsfig{file =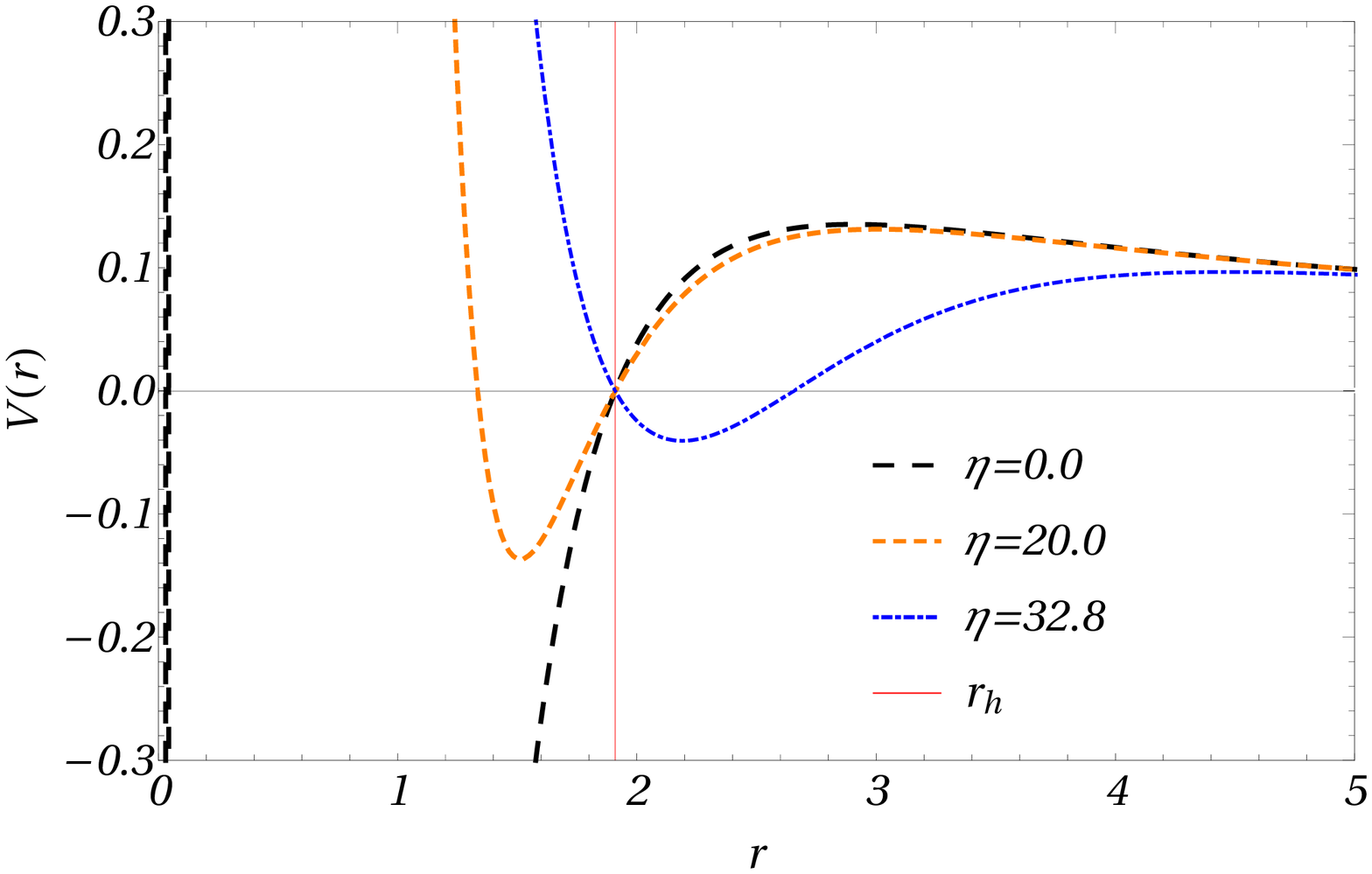, width=8.1cm, height=5.5cm}
\epsfig{file =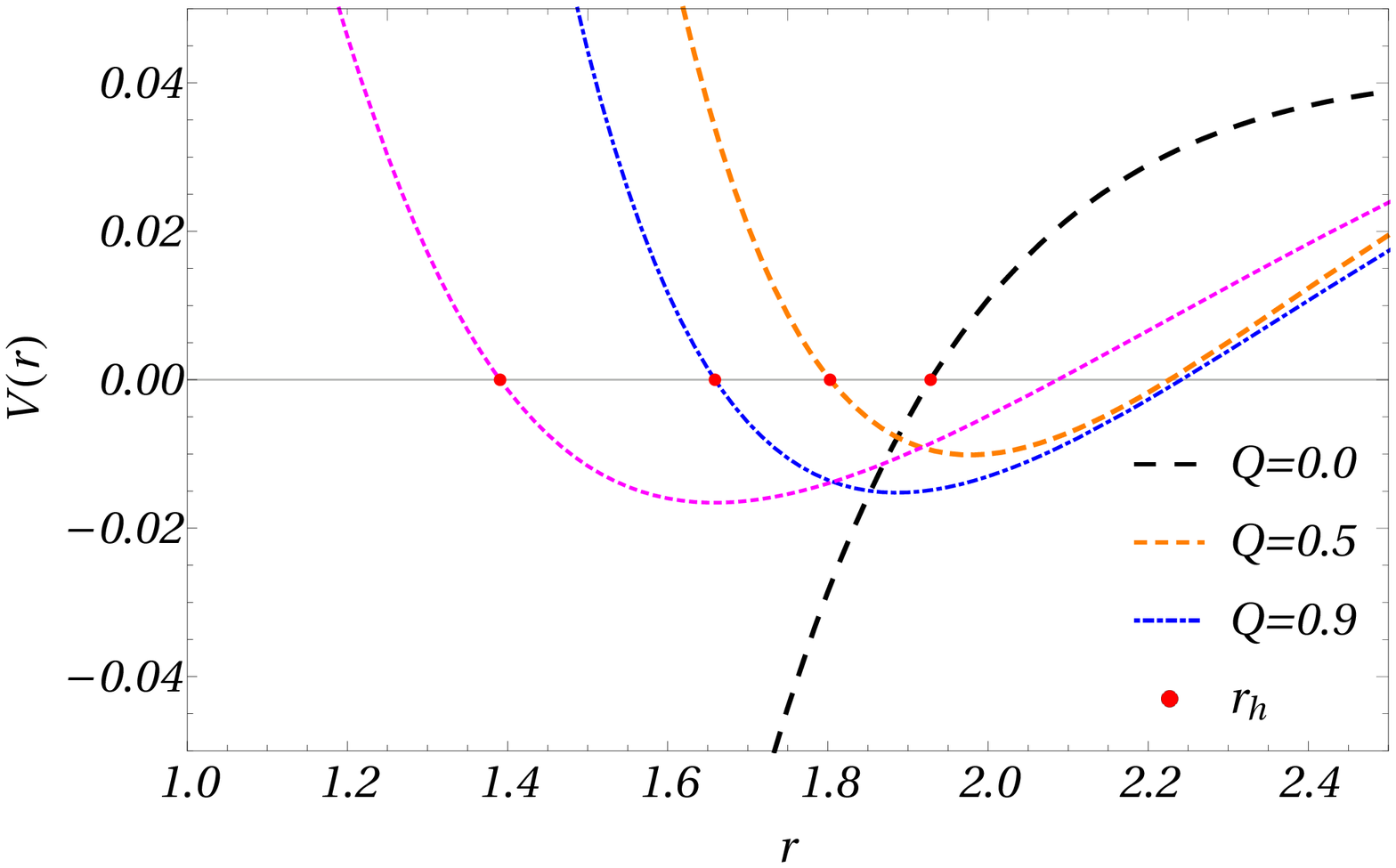, width=8.1cm, height=5.5cm}
\end{center}
\caption{{\small Effective scalar potential with parameters $M=L/10=1$ and $m=0$. Left panel: different values of $\eta$ with $l=1$ and $Q=0.2$. Right panel: different values of charge $Q$, with $l=0$ and $\eta=30$.  }}
\label{pot1}
\end{figure}

\begin{figure}[!ht]
\begin{center}
\epsfig{file =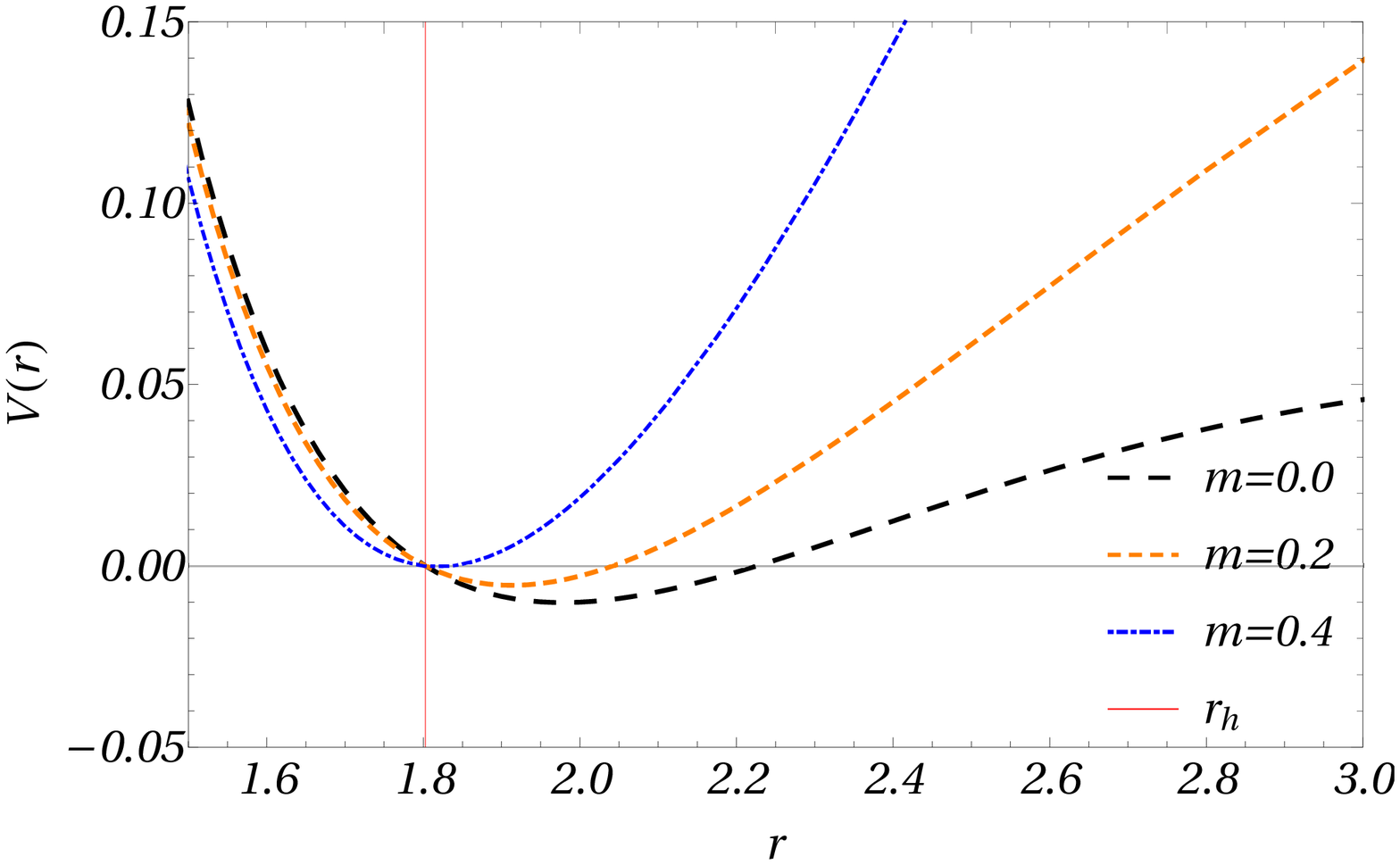, width=8.1cm, height=5.5cm}
\epsfig{file =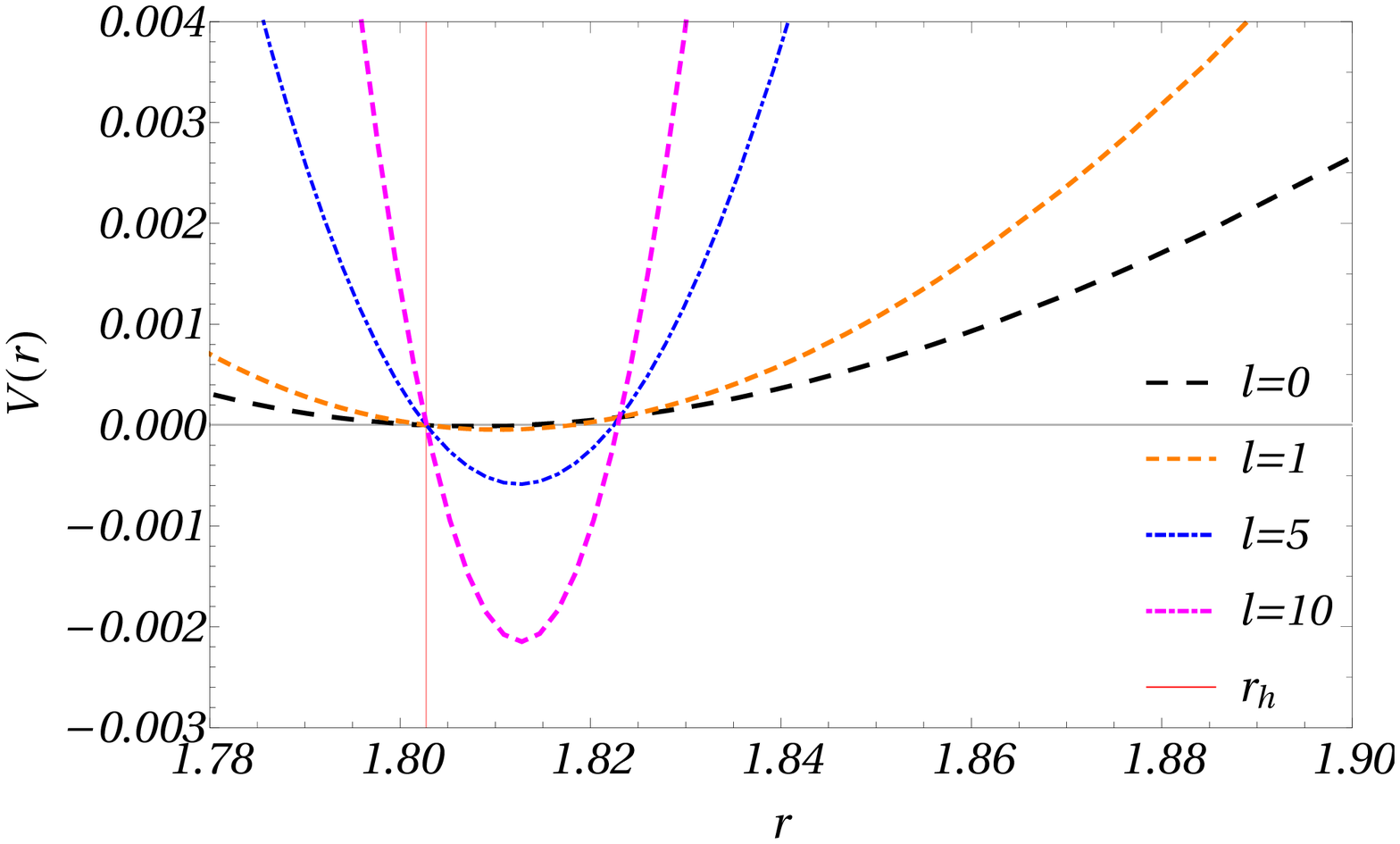, width=8.1cm, height=5.5cm}
\end{center}
\caption{{\small Effective scalar potential with parameters $M=L/10=1$. Left panel: different values of scalar field mass $m$ with $l=0$, $Q=0.5$, and $\eta=30$. Right panel: different values of multipole number $l$, $Q=0.5$, $\eta=19$, and $m=0$. }}
\label{pot2}
\end{figure}

Let us first discuss the massless perturbation case.
As can be seen in Figs. \ref{pot1} and
\ref{pot2}  regarding the
multipole number, its effect is related to the potential's peak or well
size, {\it i.e.}, as $l$ grows, the well becomes deeper and the
local peak becomes higher. However, The most interesting changes in the potential occur
when we turn on the derivative coupling $\eta$. As $\eta$ grows the
potential develops a negative well which initially stays completely inside the
event horizon for small values of $\eta$. At intermediate
values of this parameter the well is gradually shifted outside the
horizon, thus, making the potential negative in that region. In other
words $\eta$ triggers the well emergence that can eventually lead to
instabilities in the background metric.

In fact, this well becomes deeper as $\eta$ approaches a critical value, $\eta_c$, that signals the precise moment when instabilities arise. This value can be numerically computed as we will see in the next section. As for the black hole charge, it deepens
the well as long as the derivative coupling parameter $\eta$ is less
than its critical value, around which the wells attain an almost
uniform depth.  The right panel in Fig. \ref{pot1} also contains the
Schwarzschild case ($Q=0$) for reference. In addition, in all figures we show the corresponding event horizon position to indicate the region of interest. In this way we can identify potentials becoming negative at some regions after this
horizon, what can lead to possible instabilities.

Concerning the massive perturbation case, the effect of the multipole
number $l$ and the derivative coupling parameter $\eta$ remains the
same as in the massless case. In addition, the black hole charge makes
the well shift outside the event horizon more quickly. Moreover, we
observe that as the perturbation mass increases, the well gets shallow and the potential grows faster
as can be seen in Fig. \ref{pot2}.

\section{QNMs of a massive scalar field coupled to Einstein tensor}
\label{sec3}

Based on the Klein-Gordon equation displayed in Section II, we will use the well-known characteristic integration in null-coordinates method to obtain the field  propagation along with the prony method to extract the quasinormal frequencies. Both techniques were used many times in specific literature in the last years and can be found in multiple references, e. g. in \cite{Konoplya:2011qq}.

The integration procedure takes place in null-coordinates $du=dt-dr_*$
and $dv=dt+dr_*$ and the boundary condition is the usual $Z
|_{frontier} \rightarrow 0$. The complementar condition we take is the
evolution of a gaussian wave package in the $u \times t$ diagram, with which we can analyse the field profile evolution. In cases where the field evolution goes as a damped oscillation we can extract the quasinormal modes with the prony method.

In order to check the quasi-frequencies obtained we use as a second tool a Frobenius-like method, based on the expansion of the wave function around the event horizon (developed by Horowitz and Hubeny in \cite{Horowitz:1999jd}).



\subsection{Field Propagation and QNMs}

The characteristic integration in a Reissner-Nordstr\"om geometry for the Klein-Gordon field has been obtained for a multiple range of parameters. The general behavior of a quasinormal oscillation takes place for the geometry without coupling as exemplified in the cases $l=0$ and $l=1$,  and for various values of $\eta$. The results we obtained with the combination of both methods are quite similar to those seen in the literature. For large black holes and cosmological constant we obtained $\omega_{l =0} = 184.99-266.33I$ and $\omega_{l = 3} = 185.04-266.32I$ with a difference smaller than  $0.03\%$ as compared with the results given in  \cite{Wang:2000gsa}\footnote{In the reference, $\omega_{l =0} = 185.04-266.32I$ and $\omega_{l = 3} = 185.00-266.38I$}.
\begin{figure}[!ht]
\begin{center}
\epsfig{file =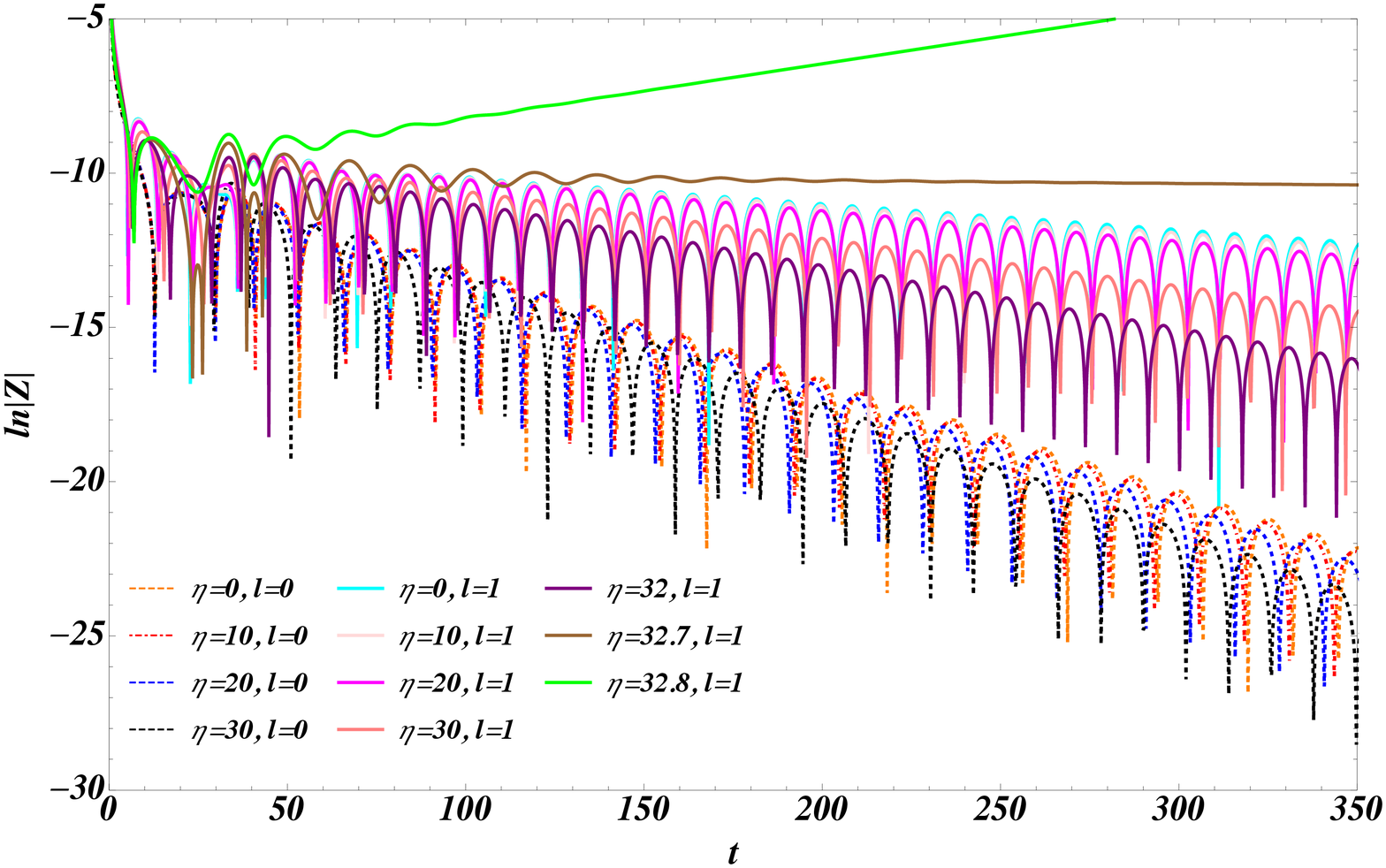, width=8.2cm, height=5.7cm}
\epsfig{file =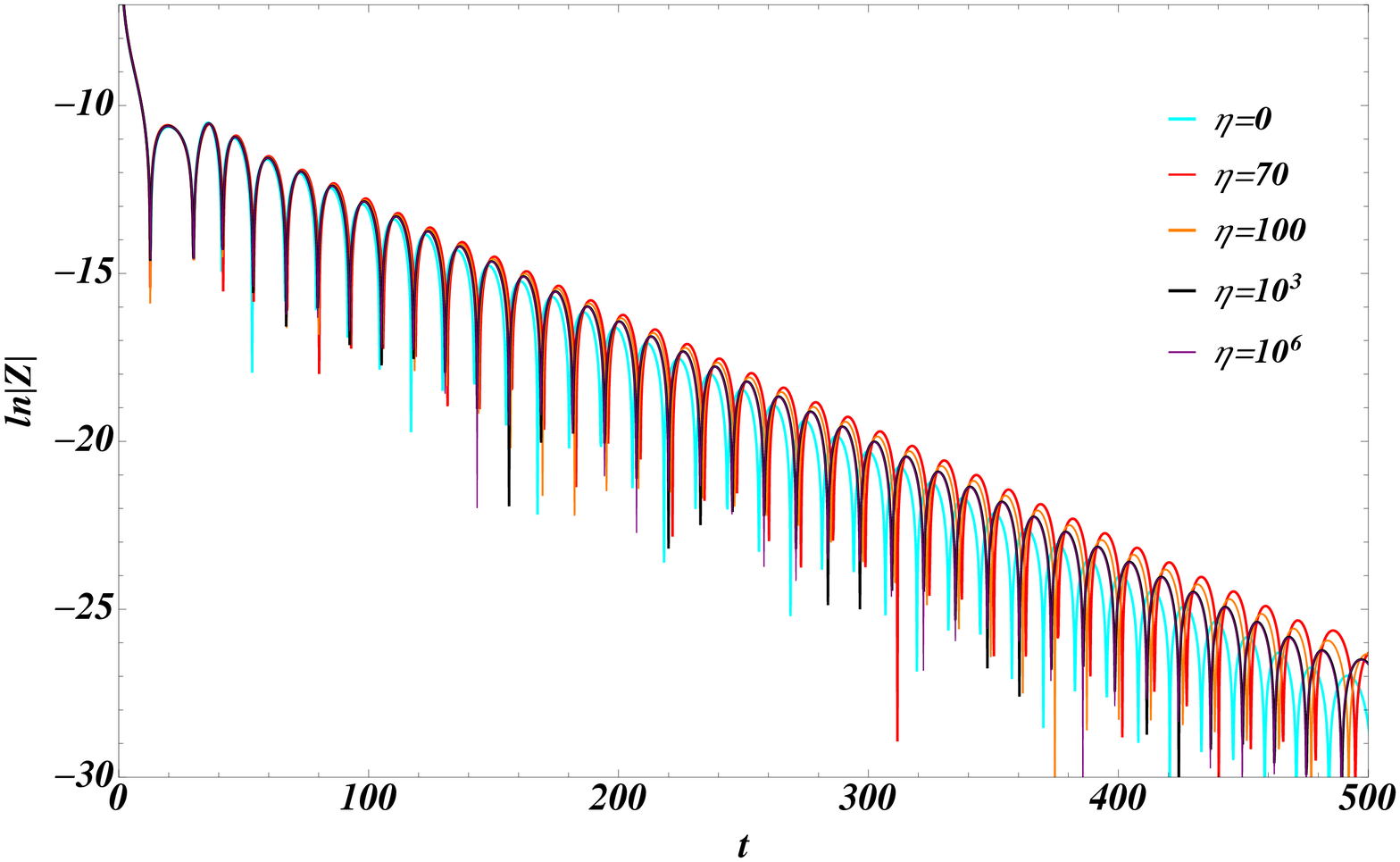, width=8.1cm, height=5.7cm}
\end{center}
\caption{{\small Scalar field behaviour for an AdS charged black hole, with varying $\eta$. The parameters of the metric are $M=L/10=5Q=1$.}}
\label{f1}
\end{figure}
In Fig. \ref{f1}, left and right panels, we see typical evolutions of
the scalar field obtained in the AdS charged geometry. The field
evolution in the massless case is stable and performing a damped oscillation profile for every $\eta$ when the wave has no angular momentum. This is also the case for other geometry parameters: the field is stable whenever $l=0$, decaying as quasinormal signal or exponentially. Otherwise, for a scalar field with $l >0$ there is always a maximum value for $\eta$ for which the evolution remains bounded. In the above-mentioned figures, for instance, if $\eta > 32.7$ (and $<100/3$) when $l>0$, the evolution will be unstable. In this case the geometry of the spacetime is expected to evolve as well and such a change has to be investigated with the full non-linear Einstein equations, what is beyond the scope of this work.
\begin{figure}[!h]
\begin{center}
\epsfig{file =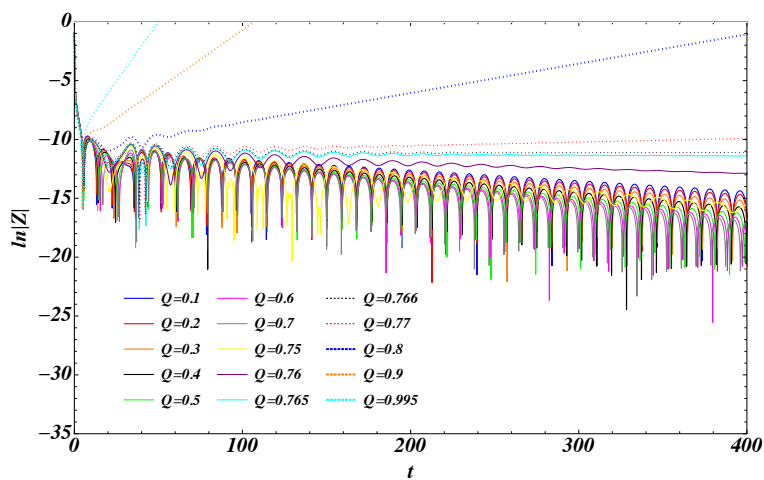, width=10cm, height=6.2cm}
\end{center}
\caption{{\small Scalar field behaviour for an AdS charged black hole with different charges. The parameters of the metric are $M=L/10=\eta /20=1$. There is a critical value of $Q$ from which the field destabilizes.}}
\label{fa2}
\end{figure}
For high values of $\eta$ the evolution of the scalar field is almost the same as we vary $\eta$, what we can see in Fig. \ref{f1}, right panel. Moreover, the quasinormal modes remain unaffected in this case, i.e., the coupling does not influence the spectra of the black hole, as we  show in the next subsection.

In Fig. \ref{fa2}, we can see typical evolutions of the scalar field profiles in charged black holes. They are qualitatively the same evolution obtained in the Reissner-Nordstr\"om-AdS case ($\eta = 0$), except near (and after) a threshold charge. In that case, $Q \sim 0.765$. Whenever $Q < 0.766$ the field evolves stably, first with a ring-down signal and for charges near the critical point, as an exponential decay. On the other hand, if $Q>0.766$, the field destabilizes and the geometry must change.

The critical value of $\eta$ for which the scalar field is not stable depends on the parameters of the geometry, as expected, and notably on the angular momentum of the field.
\begin{table}[htbp]
\centering
 \begin{tabular}{cccccccc}
    \hline
      \multicolumn{8}{c}{$10M=L=10$} \\
	\hline
     $l$ & $\hspace{0.2cm} $ $Q=0.1$  $\hspace{0.2cm} $ & $\hspace{0.2cm} $ $Q=0.2$ $\hspace{0.2cm} $ & $\hspace{0.2cm} $ $Q=0.4$ $\hspace{0.2cm} $ & $\hspace{0.2cm} $ $Q=0.6$ $\hspace{0.2cm} $ & $\hspace{0.2cm} $ $Q=0.8$ $\hspace{0.2cm} $ & $\hspace{0.2cm} $ $Q=0.95$ $\hspace{0.2cm} $  & $\hspace{0.2cm} $ $Q=0.99517$\\
      \hline
    1	& $33.15^{^{^ {\pm 0.05}}}$ & $32.75^{^{^ {\pm 0.05}}}$ & $30.65^{^{^ {\pm 0.05}}}$ & $26.35^{^{^ {\pm 0.05}}}$ & $18.25^{^{^ {\pm 0.05}}}$ & $7.95^{^{^ {\pm 0.05}}}$ & $1.75^{^{^ {\pm 0.05}}}$ 	\\
    2	& $33.05^{^{^ {\pm 0.05}}}$ & $32.15^{^{^ {\pm 0.05}}}$ & $28.65^{^{^ {\pm 0.05}}}$ & $22.35^{^{^ {\pm 0.05}}}$ & $13.45^{^{^ {\pm 0.05}}}$ & $5.25^{^{^ {\pm 0.05}}}$ & $1.25^{^{^ {\pm 0.05}}}$ 	\\
    3	& $32.95^{^{^ {\pm 0.05}}}$ & $31.85^{^{^ {\pm 0.05}}}$ & $27.45^{^{^ {\pm 0.05}}}$ & $20.35^{^{^ {\pm 0.05}}}$ & $11.55^{^{^ {\pm 0.05}}}$ & $4.35^{^{^ {\pm 0.05}}}$  & $1.15^{^{^ {\pm 0.05}}}$ 	\\
    \hline
    \hline
    \multicolumn{8}{c}{Lowest limit of stability for large $l$ using Eq.(\ref{etalim})} \\
    \hline
   $\infty$ &32.54 &30.29 &22.91&14.44&7.10&2.63&0.92\\
    \hline
  \end{tabular}
   \caption{Critical value of $\eta$ for the scalar field for
     different charges of the geometry (in unities of $Q_{ext}$) and
     angular momentum of the field. For the geometry parameters,
     $Q_{ext} \sim 0.99518$. The corresponding values of $\eta_{lim}$
     are also shown for reference.}
\label{etac1}
\end{table}

In Table \ref{etac1} we list some of these values for the
Reissner-Nordstr\"om-AdS black hole. We observe that the higher the value of the
charge in the geometry is, the smaller the value of critical $\eta$
will be, the same being true for other multipole numbers. The
transitional value  of $\eta$ in relation to the stability of the
scalar field achieves the highest gap from $l=1$ to $l=3$ around
$Q=0.8Q_{ext}$ and decreases at the extremal charge. For example, if
$Q=0.1Q_{ext}$, the value of transition in $\eta$ varies slowly from
$l=1$ to $l=3$ (circa $0.6\%$), while for $Q=0.8Q_{ext}$, from $l=1$
to $l=3$ the variation for $\eta$ increases to $36.7\%$ and for
$Q=0.99999Q_{ext}$, $34.3\%$. This means that accreting charge in a
Schwarzschild black hole (with a higher rate than the accretion of
mass) causes the reduction of the range of stability in $\eta$ for
which the scalar field evolution decays in time.

\begin{figure}[!ht]
\begin{center}
\epsfig{file =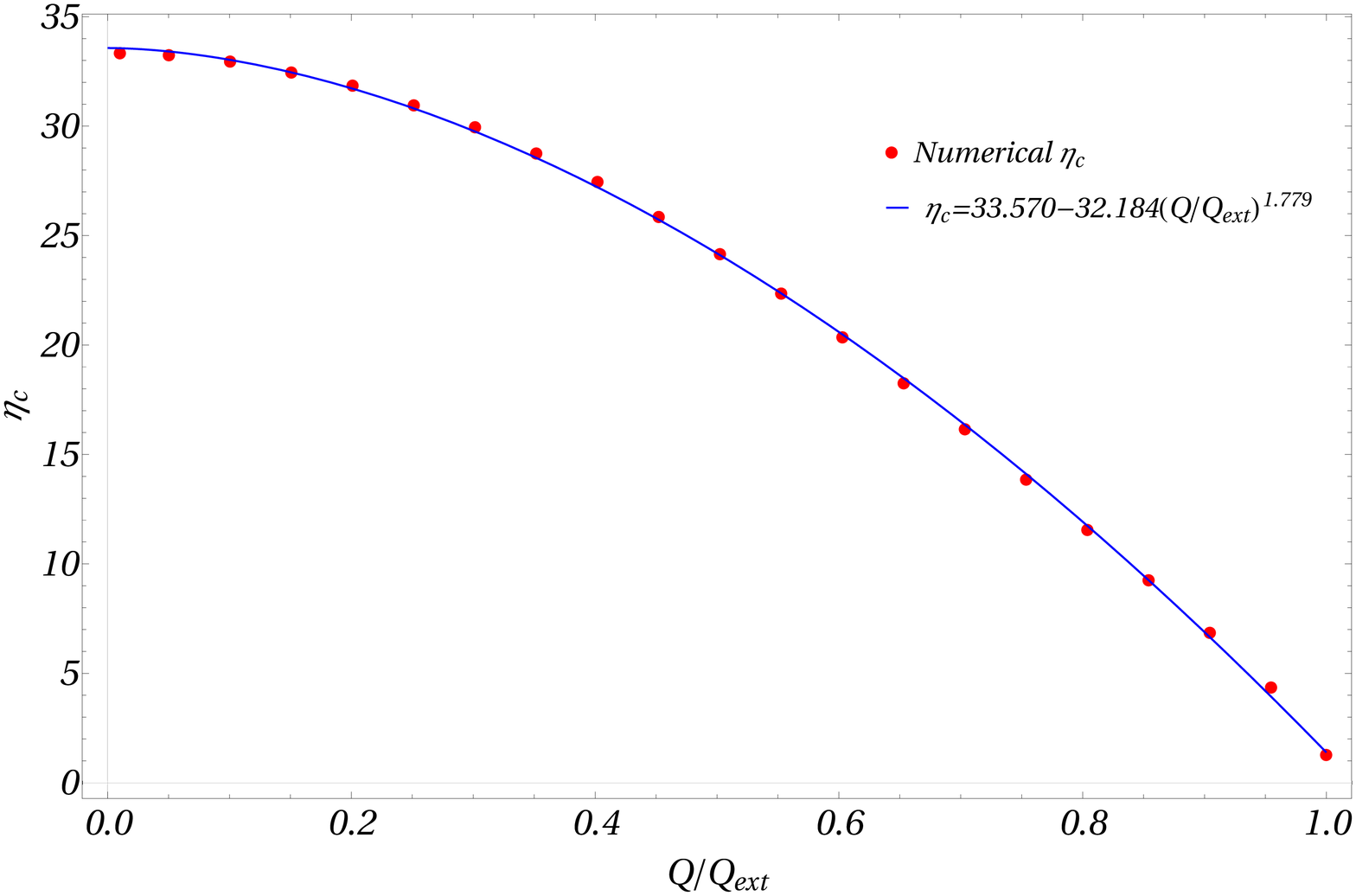, width=10cm, height=6.2cm}
\end{center}
\caption{{\small Critical value of $\eta$ as a function of the charge
    (in units of $Q_{ext}$) in  Reissner-Nordstr\"om-AdS black
    hole. The parameters of the metric are $M=L/10=1$ and for the
    field $m=0$ and $l=3$.}}
\label{f2}
\end{figure}

In Fig. \ref{f2}
we plotted our results for critical values of $\eta$ as a function of the
black hole charge (in units of $Q_{ext}$). We also show the corresponding
fitting chosen to be the simplest function of a power of the charge with three
parameters,
\begin{equation}\label{fit1}
\eta_c = 33.570-32.184\, (Q/Q_{ext})^{1.779}\,.
\end{equation}
This fitting
produces a factor $R^2=0.99994$, where we define
\begin{equation}
R^2\equiv 1-\frac{\sum\limits_i (y_i-f_i)^2}{\sum\limits_i (y_i-\bar y)^2} \,,
\end{equation}
being $y_i$ and $\bar y$ the data we numerically calculated and
their corresponding mean value and $f_i$, the value produced by the
fitting function. Clearly, the value $R^2=1$ means a perfect fitting.
This shows the excellent correlation with
the points numerically calculated.

The instability of the scalar field increases for increasing $\eta$ up
to $L^2 /3$. After this value the field transformation we used
generates a discontinuity in the potential for $r>r_h$, and we shall
not study the region of parameters for which $r_d>r_h$ until $r_d$ for
high enough $\eta$ becomes imaginary. At this point the discontinuity disappears and the integration of the scalar field equation produces only stable evolutions, in as much as the potential is positive definite again.

In order to gain some insight about the critical value of $\eta$ let
us analyse the behaviour of the effective potential near the
horizon. In this region it can be rewritten as $V_{r_h}\sim F(r)\ \Omega(r_h)$ where
\begin{eqnarray}
\label{Vnearhorizon}
\Omega(r_h)&=&\frac{m^2 r_h^2+l(l+1) \left[1-\left(\frac{3}{L^2}+\frac{Q^2}{r_h^4}\right) \eta \right]}{r_h^2\left[1+\left(-\frac{3}{L^2}+\frac{Q^2}{r_h^4}\right)\eta \right] }+\frac{F'(r_h)}{r_h}-\frac{2 Q^2 \eta  F'(r_h)}{r_h^5 \left[1+\left(-\frac{3}{L^2}+\frac{Q^2}{r_h^4}\right) \eta \right]}
\end{eqnarray}
Since $F(r)$ is a positive function, the change of sign in the
potential necessarily comes from $\Omega(r_h)$. In this way, if the potential
turns to be negative at some regions, unstable modes could in
principle be turned on. Thus, we search for the zeros of $\Omega(r_h)$
and find the value of $\eta$ when this change of sign happens,
\begin{eqnarray}
\label{eta0}
\eta_0\sim \frac{-3 L^2 r_h^8+L^4 r_h^4 \left[Q^2-r_h^2 \left(1+l(l+1)+m^2 r_h^2\right)\right]}{\left(L^2 Q^2+3 r_h^4\right) \left\{-3 r_h^4+L^2 \left[Q^2-r_h^2(1+l(l+1)) \right]\right\}}.
\end{eqnarray}
The first thing we notice is that when $m=0$, Eq.(\ref{eta0})
becomes independent of $l$ and is reduced to the limit value,
\begin{eqnarray}
\label{etalim}
\eta_{lim}\sim \frac{L^2 r_h^4}{L^2 Q^2+3 r_h^4}
\end{eqnarray}
This same result also corresponds to the limit when
$l\rightarrow\infty$, {\it i.e.}, for large multipole numbers.
This independence can be seen in the right panel of
Fig. \ref{pot2}. For large $l$ the zeros of the potential remain at the
same position and $l$ only affects the well depth. Moreover, our
calculations show that as $l$ increases, $\eta_c$ approaches
$\eta_{lim}$ as we can see in Table \ref{etac1}. For instance, the
numerical results for large $l$ resemble the approximation above
listed in table I. If we take $M=0.1L=5Q=l/50=1$, then  $\eta_c \sim 30.45$, only $0.5\%$ away from the listed value for $l \rightarrow \infty$.
Thus, $\eta_{lim}$ can be considered the lowest limit of stability for
large $l$.

Inspired by Eq.(\ref{eta0}) we also found an alternative fitting for
the numerical data in Fig. \ref{f2} given by
\begin{equation}\label{fit4}
\eta_c = \frac{8.103-7.617\,(Q/Q_{ext})^2}{\left[0.518+0.004\,(Q/Q_{ext})\right]\left[0.469+0.091\,(Q/Q_{ext})^2\right]}\,,
\end{equation}
which has a similar $R^2$ factor as Eq.(\ref{fit1}) showing excellent agreement with the
numerical data.

\begin{figure}[!ht]
\begin{center}
\epsfig{file =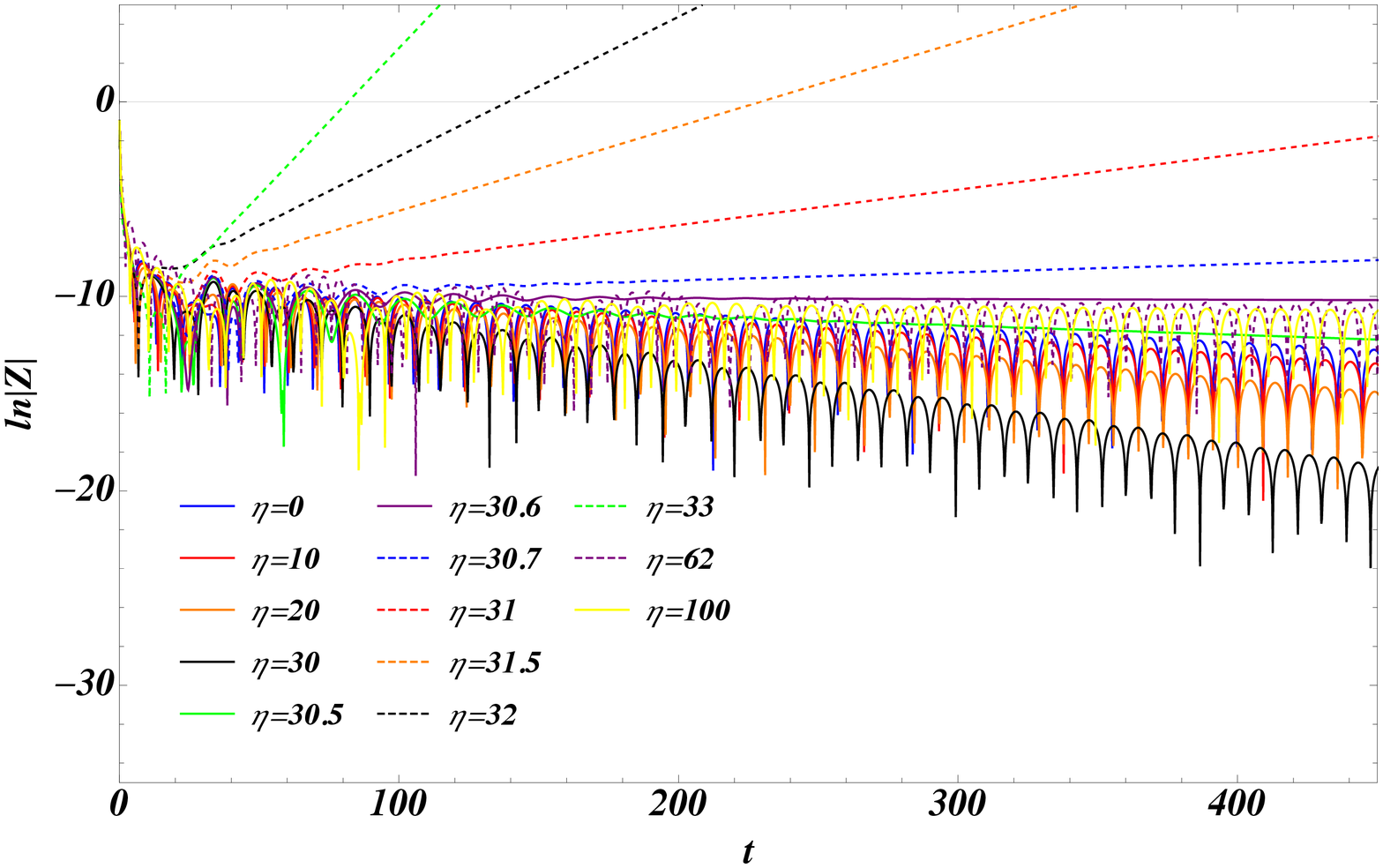, width=8.1cm, height=5.5cm}
\epsfig{file =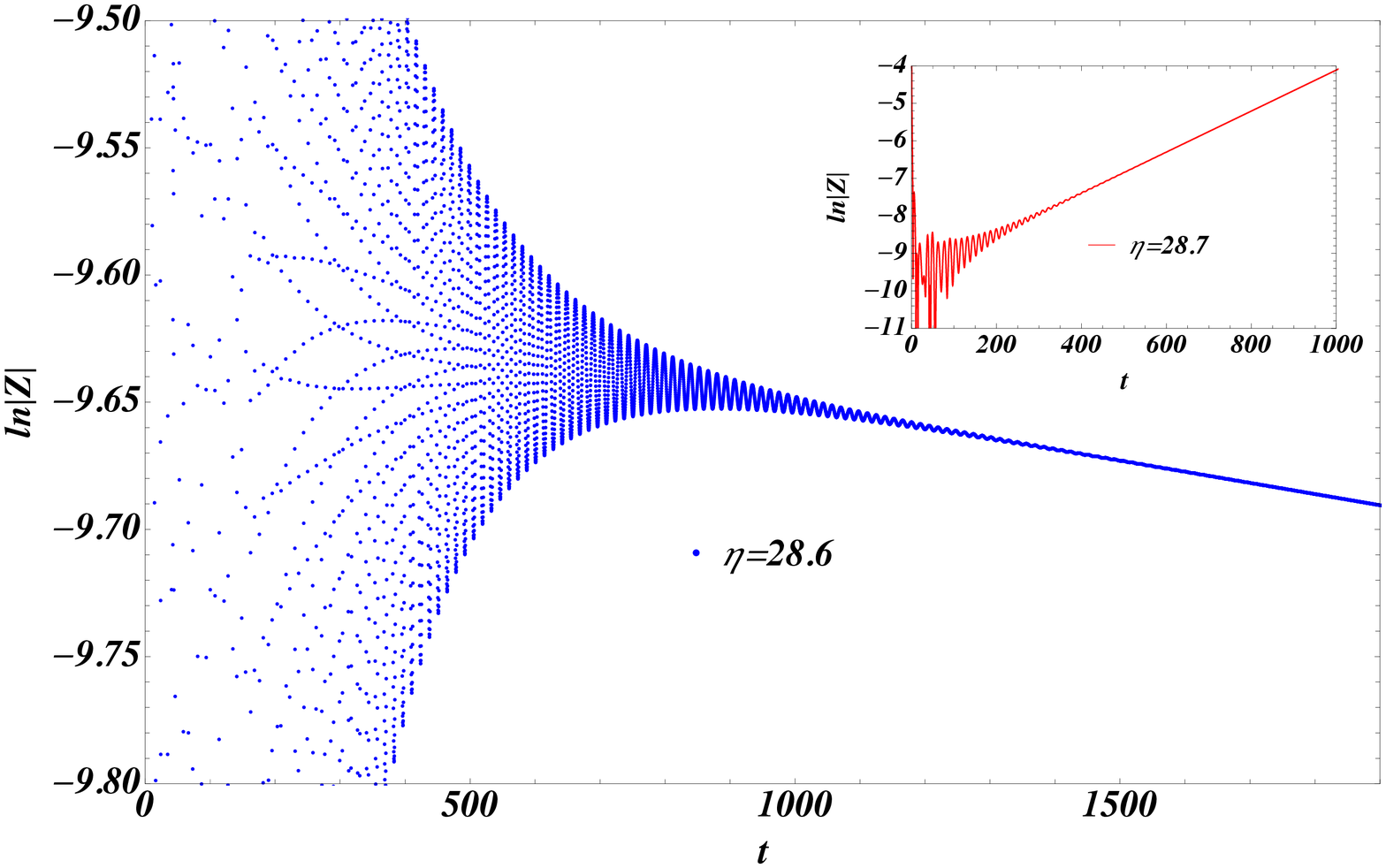, width=8.1cm, height=5.5cm}
\end{center}
\caption{{\small Scalar field evolution for multiple values of $\eta$ on the charged AdS black hole. The parameters of the metric are $M=L/10=2.5Q=1$ and for the field $l=1$ (left) and $l=2$ (right). The threshold of stability for $\eta$ is different for each $l$. In the left panel we see a critical coupling constant $\eta_c \sim 30.65$, and in the right panel $\eta_c \sim 28.65$ (the blue points presenting a stable evolution and the red one unstable).}}
\label{f3}
\end{figure}
In Fig. \ref{f3} we can see examples of this limiar of stability, for $l =1$ and $l=2$. In the first case $\eta_c \sim 30.6$ and for $\eta > 62$ only stable evolutions are seen for every $l$. The same behavior is seen in the right panel for $l =2$ in which $\eta_c \sim 28.6$, and again for $\eta > 62$ only stable profiles are generated. In cases of very high $\eta$ no difference is noticed in the field evolution as shown in Fig. \ref{f1}: for $\eta= 10^3$ to $10^{40}$ all signals collapse to a single one.

The situation is qualitatively similar when we study the coupling for large black holes and large $\Lambda$. In such a case it is not possible to verify qualitative changes in the quasinormal spectra from that of an AdS black hole without coupling and the quasinormal modes are marginally affected except when $\eta \sim \frac{L^2}{3}$ (we compute some examples in the next subsection). This comes with no surprise if we look at Table \ref{etac1} and Fig. \ref{f3}: whenever $\eta > \eta_c$, the field destabilizes more and more as $\eta$ approaches $\frac{L^2}{3}$. Thus, for black holes with high $r_h$, a discerning influence is generated in the field propagation only for $\eta \sim \Lambda^{-1}$. As we may further see, the quasinormal modes do not change in such cases.

\subsection{The quasinormal frequencies}

The quasinormal frequencies obtained for the Reissner-Nordstr\"om-AdS black hole are mostly affected by the scalar field coupling in the limit of small $r_h$ and $L$. In the cases of high $r_h$ or $\eta >> \Lambda^{-1}$, the effect of the coupling is very mild.
\begin{table}[htbp]
  \centering
 \caption{The quasinormal modes of the RNAdS black hole with $l =0$.}
    \begin{tabular}{cccccc}
    \hline
 \multicolumn{3}{c}{$M=L/10=5Q=1$} \hspace{1.0cm}&  \multicolumn{3}{c}{$r_h=5Q=50L=50$} \\
	\hline
    $\eta$  & {$\Re(\omega)$} & {$-\Im(\omega)$} &  $\eta$ & {$\Re(\omega)$} & {$-\Im(\omega)$}  \\
\hline \hline
    70 & 0.2443 & 0.03371 &      0     & 92.548 & 133.17 \\
    100 & 0.2452 & 0.03437 &     0.3 & 92.551 & 133.18 \\
    1000 & 0.2462 & 0.03507 &    0.33 & 92.579 & 133.30 \\
    $10^6$ & 0.2463 & 0.03511 &  0.333 & 92.891 & 134.50 \\
    $10^{12}$ & 0.2463 & 0.03511 & 0.3333 & 100.20 & 146.53 \\
    $10^{40}$ & 0.2463 & 0.03511 & 1 & 92.548 & 133.17 \\
   \hline
    \end{tabular}
  \label{t2}
\end{table}
In Table \ref{t2},  the frequencies vary less than $0.2\%$ for every $ \eta > 30\Lambda^{-1}$ for small black holes. On the other hand, in the limit of high $r_h$, the scalar field spectrum remains unaffected for the coupling except very near $\Lambda^{-1}$: the quasinormal mode is exactly the same (to the fifth figure) for $\eta =0$ and every $\eta >1$. A similar behavior is obtained varying the parameters, maintaining $r_h$ high: if we take, e. g., $r_h=Q=100L=100$ and $l=0$, the fundamental mode is $\omega = 184.95-266.38I$ for $\eta = 0$ and
$\omega = 184.95-266.36I$ for every $\eta \geq 1$, which also occurs for other values of $l$\footnote{For  $l=2$, $\omega = 184.98-266.37I$ for $\eta = 0$ and $\omega = 184.98-266.35I$ for every $\eta \geq 1$.}.

\begin{table}[htbp]
  \centering
 \caption{The quasinormal modes of the RNAdS black hole with $M=L/10=1$ and $l =0$.}
    \begin{tabular}{ccccccccccccc}
    \hline
	\multirow{3}{*}{$\eta$} & \multicolumn{2}{c}{$Q=0.1$} & \multicolumn{2}{c}{$Q=0.2$} & \multicolumn{2}{c}{$Q=0.4$} & \multicolumn{2}{c}{$Q=0.6$} & \multicolumn{2}{c}{$Q=0.8$} & \multicolumn{2}{c}{$Q=0.95$} \\
	\hline
& {$\Re(\omega)$} & {$-\Im(\omega)$} & {$\Re(\omega)$} & {$-\Im(\omega)$} & {$\Re(\omega)$} & {$-\Im(\omega)$} & {$\Re(\omega)$} & {$-\Im(\omega)$} & {$\Re(\omega)$} & {$-\Im(\omega)$} & {$\Re(\omega)$} & {$-\Im(\omega)$} \\
	\hline
    0     & 0.2453 & 0.03664 & 0.2483 & 0.03641 & 0.2477 & 0.03542 & 0.2464 & 0.03372 & 0.2440 & 0.03125 & 0.2407 & 0.02962 \\
    5     & 0.2486 & 0.03669 & 0.2487 & 0.03661 & 0.2492 & 0.03625 & 0.2505 & 0.03551 & 0.2533 & 0.03363 & 0.2559 & 0.02959 \\
    10    & 0.2487 & 0.03676 & 0.2492 & 0.03689 & 0.2514 & 0.03721 & 0.2556 & 0.03688 & 0.2614 & 0.03395 & 0.2645 & 0.02905 \\
    15    & 0.2489 & 0.03687 & 0.2500 & 0.03730 & 0.2545 & 0.03830 & 0.2616 & 0.03760 & 0.2687 & 0.03347 & 0.2716 & 0.02880 \\
    20    & 0.2493 & 0.03706 & 0.2513 & 0.03794 & 0.2592 & 0.03938 & 0.2688 & 0.03754 & 0.2761 & 0.03285 & 0.2786 & 0.02896 \\
    25    & 0.2500 & 0.03746 & 0.2541 & 0.03905 & 0.2669 & 0.03997 & 0.2779 & 0.03673 & 0.2847 & 0.03246 & 0.2872 & 0.02966 \\
    30    & 0.2528 & 0.03879 & 0.2632 & 0.04101 & 0.2818 & 0.03886 & 0.2925 & 0.03590 & 0.2995 & 0.03423 & ***   & *** \\
	\hline
    \end{tabular}
  \label{t3}
\end{table}
In Table \ref{t3} we can see the effect of $\eta$ coupling in the quasinormal spectra for a small black hole. The influence is again more pronounced in the regions $\eta \sim \Lambda^{-1}$ specially for high values of charge. In general the field profile rapidly undergoes the exponential decay for $\eta$ near $\Lambda^{-1}$. For example, no oscillation forms for $Q=0.95$ and $\eta=30$. The oscillation of the values of the quasinormal modes (increasing and decreasing with increasing charge) is an expected feature already demonstrated in other references \cite{Wang:2004bv}.

In addition, quasinormal frequencies were calculated using another
numerical approach, developed by Horowitz and Hubeny
\cite{Horowitz:1999jd}, as a double check. As is usually the case,
this method produces the best results for large $r_h$, while for small
$r_h$ the convergence of the method is problematic. The values we
found are in good agreement with the previous method and the
difference between these values is around $0.05\%$ for the real part
and $0.03\%$ for the imaginary part of the frequencies when $\eta$ is
far from the critical value. Near $\eta_c$ the convergence of
Horowitz-Hubeny method shows to be very poor.

\section{Final remarks}\label{sec5}

In the present work we have investigated the influence of a non-minimal derivative coupling $\eta$ of a massive scalar field coupled to Einstein tensor on the propagation of this field in the vicinity of a Reissner-Nordstr\"om-AdS black hole. We carried out a detailed investigation of the  regions of instability of the background black hole which arise depending upon the value of  $\eta$ and the parameters of the theory, namely the mass $M$ and the electric charge $Q$ of the black hole, the AdS radius $L$ and also the scalar field multipole number $l$ and its mass $m$.

In the case of massless scalar perturbations the effective potential develops a negative well, which can be shifted from inside the event horizon to the exterior region as the derivative coupling parameter $\eta$ grows and can be made deep enough depending on the region of parameters. The development of a negative well indicates possible instabilities of the background Reissner-Nordstr\"om-AdS black hole and it is confirmed by the analysis of field propagation through the computation of quasinormal modes and frequencies. In the case of massive scalar perturbations  as the perturbation mass increases, the well gets shallow and the potential grows faster.

In the case of zero angular momentum the massless scalar field evolves stably
whatever the value of $\eta$ is. However, for non-zero angular momentum we found a critical value of the derivative coupling $\eta_c$ above which  the scalar field propagates unstable modes. Looking at the QNMs we observed that as we
increase $\eta$ above its critical value  $\eta_c$, the oscillations are getting slower and the QNMs decay faster. This behaviour is expected because as we already discussed, the coupling of the scalar field to Einstein tensor strongly influences its kinetic energy.

Regarding the effect of the black hole charge $Q$, we found that as the charge $Q$ is approaching its extremal value $Q_{ext}$, the critical value $\eta_c$ is decreasing.  A similar behaviour was observed in a charged rotating black hole. It was found that instabilities can appear  when the angular momentum of the
black hole is small, as long as the charge is sufficiently large \cite{Andrade:2018rcx,Tanabe:2016opw}.

Finally, as we already discussed, for values of $\eta$ beyond $\eta_c$ the field develop instabilities. However, we observed that stability is recovered after a certain value of $\eta > \eta_c$, featuring two transitions of the scalar field propagation from stability to instability and going back to stable quasinormal oscillation. This is an intriguing effect indicating a kind of phase transition from an unstable to a stable configuration, observed also in de Sitter geometries \cite{EtadS}.

This behaviour may signals that the Reissner-Nordstr\"om-AdS black hole is scalarized, i.e., it acquires hair and it gets stabilized. Actually a similar behaviour was found in \cite{Doneva:2018rou} in which the extended scalar-tensor-Gauss-Bonnet gravity was studied and it was found that a scalar field, sourced by the curvature of the spacetime via the Gauss-Bonnet invariant, scalarized spontaneously the Reissner-Nordstr\"om-AdS black hole. We intend to further study this effect in a fully backreacting problem with the scalar field interacting with the background metric in a future project.

\section{Acknowledgments}

 This work was supported by CNPq (Conselho Nacional de Desenvolvimento Cient\'{\i}fico e Tecnol\'ogico), FAPESP (Funda\c c\~ao de Amparo \`a Pesquisa do Estado de S\~ao Paulo), and FAPEMIG (Funda\c c\~ao de Amparo \`a Pesquisa do Estado de Minas Gerais), Brazil. E.P.  acknowledges the hospitality of the Physics Institute of the University of S\~ao Paulo where this work started and CNPq for financial support.


\end{document}